\documentclass[twocolumn,aps,prl,amsmath,amssymb,showpacs,superscriptaddress]{revtex4}

\usepackage{graphicx,color}
\usepackage{dcolumn}
\usepackage{epsfig}
\usepackage{bm}

\newcommand{\ket}[1]{\ensuremath{|\,{#1}\,\rangle}}
\newcommand{\bra}[1]{\ensuremath{\langle\,{#1}\,|}}

\begin{document}

\title{Experimental Bell inequality violation without the postselection loophole}

\author{G. Lima}
%\email{glima@udec.cl}
\affiliation{Center for Optics and Photonics, Universidad de
Concepci\'{o}n, Casilla 4016, Concepci\'{o}n, Chile.}

\author{G. Vallone}
%\email{giuseppe.vallone@uniroma1.it}
\affiliation{Museo Storico della Fisica e Centro Studi e Ricerche
``Enrico Fermi'', Via Panisperna 89/A, Compendio del Viminale, Roma
I-00184, Italy}
\affiliation{Dipartimento di Fisica, Universit\`{a} Sapienza di Roma, I-00185 Roma, Italy}

\author{A. Chiuri}
\affiliation{Dipartimento di Fisica, Universit\`{a} Sapienza di Roma, I-00185 Roma, Italy}

\author{A. Cabello}
% \email{adan@us.es}
 \affiliation{Departamento de F\'{i}sica Aplicada II, Universidad de Sevilla, E-41012 Sevilla, Spain}

\author{P. Mataloni}
\email{paolo.mataloni@uniroma1.it}
\affiliation{Dipartimento di Fisica, Universit\`{a} Sapienza di Roma, I-00185 Roma, Italy}
\affiliation{Istituto Nazionale di Ottica Applicata (INOA-CNR), L.go
E. Fermi 6, 50125 Florence, Italy}

%%%%%%%%%%%%%%%%%%%%%%%%%%%%%%%%%%%%%%%%%%%%%%%%%%%%%%%%%%%%%%%%%%%

\begin{abstract}
We report on an experimental violation of the
Bell-Clauser-Horne-Shimony-Holt (Bell-CHSH) inequality using
energy-time entangled photons. The experiment is not free of
the locality and detection loopholes, but is the first
violation of the Bell-CHSH inequality using energy-time 
entangled photons which is free of the postselection
loophole described by Aerts {\em et al.} [Phys. Rev. Lett. {\bf
83}, 2872 (1999)].
\end{abstract}

%%%%%%%%%%%%%%%%%%%%%%%%%%%%%%%%%%%%%%%%%%%%%%%%%%%%%%%%%%%%%%%%%%%

\pacs{03.65.Ud,03.67.Mn,42.65.Lm,07.60.Ly}

%03.65.Ud: Entanglement and quantum nonlocality
%(e.g. EPR paradox, Bell's inequalities, GHZ states, etc.)
%03.67.Mn Entanglement production, characterization, and manipulation
%42.65.Lm: Parametric down conversion and production of entangled photons
%07.60.Ly: Interferometers

\maketitle

%%%%%%%%%%%%%%%%%%%%%%%%%%%%%%%%%%%%%%%%%%%%%%%%%%%%%%%%%%%%%%%%%%%

{\em Introduction.---}The violation of Bell inequalities
\cite{bell64phy, clau69prl} has both profound implications for
our understanding of the universe, and striking applications
like eavesdropping-proof communication \cite{barr05prl},
reduction of communication complexity \cite{buhr09rmp}, and
randomness certification \cite{piro09nat}. Testing Bell
inequalities requires reliable methods for entangling,
distributing, and measuring particles in spacially separated
regions. So far, no experiment \cite{free72prl, aspe82prl,
weih98prl, rowe01nat, mats08prl} have shown a conclusive
violation of a Bell inequality. Local hidden variable (HV)
models exist which reproduce the results of any performed
experiment. These local HV models exploit the locality and the
detection loopholes \cite{aspe99nat, sant91prl}.

In 1989, Franson \cite{fran89prl} introduced a setup which
allows to create and measure two energy-time entangled photons
over large distances. Franson's setup used the essential
uncertainty in the time of emission of a pair of photons to
make undistinguishable two alternative paths that the photons
can take. As a result, photons detected in coincidence become
entangled in path. This type of entanglement is usually called
``energy-time'' or ``time-bin'' entanglement, depending on the
method used to have uncertainty in the time of emission
\cite{bren99prl}. Franson's setup is widely used for testing
violations of Bell inequalities \cite{kwia90pra, ouzo90prl,
bren91prl, kwia93pra, taps94prl, titt98prl, titt00prl,
rib00pra, thew04prl, marc04prl, sala08prl, sala08nat}.

However, in the Franson's scheme a new loophole, the ``postselection'' loophole,
is present. Indeed, Aerts {\em et al.} \cite{aert01prl} showed that, even
in the ideal case of two-particle preparation, enough spatial
separation between the local measurements (which closes the locality loophole), and perfect
detection efficiency (which closes the detection loophole), there are local HV models that reproduce
the quantum predictions for the violation of the
Bell-Clauser-Horne-Shimony-Holt (Bell-CHSH) inequality
\cite{clau69prl} using Franson's setup. This is because, in
Franson's setup, the fact that photons are detected in
coincidence or not could depend on the local measurement
settings. This can be exploited to build local HV models which
simulate the quantum predictions for this experiment
\cite{aert01prl, cabe09prl}.

%%%%%%%%%%%%%%%%%%%%%%%%%%%%%%%%%%%%%%%%%%%%%%%%%%%%%%%%%%%%%%%%%%%
% Fig. 1
%%%%%%%%%%%%%%%%%%%%%%%%%%%%%%%%%%%%%%%%%%%%%%%%%%%%%%%%%%%%%%%%%%%

\begin{figure}[t]
\centering
\includegraphics[width=8.2cm]{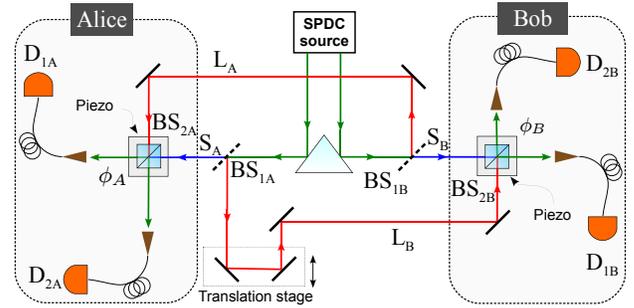}
\caption{Experimental setup. A beta barium borate crystal splits incoming photons
into pairs of photons of lower energy in a spontaneous parametric down conversion (SPDC) process.
The time of emission of each pair is unpredictable.
A step-by-step translation stage allows us to create
the indistinguishability condition ($L_A-S_A=L_B-S_B$).
Detectors' dead time is 1 ns and, due to continuous wave pumping,
the probability of two photon pair events is negligible.
Alice and Bob's spacially separated experiments are indicated by boxes.
The distance between Alice and Bob's experiments is approximately one meter.
The local phases $\phi_A$ and $\phi_B$ can be finely set by independent piezoelectric
adjustment of beam splitters $BS_{2A}$ and $BS_{2B}$, respectively.}
\label{Fig:Setup}
\end{figure}

%%%%%%%%%%%%%%%%%%%%%%%%%%%%%%%%%%%%%%%%%%%%%%%%%%%%%%%%%%%%%%%%%%%

One way to deal with this ``postselection loophole'' that
affects to {\em all} performed Bell tests using energy-time or
time-bin entangled photons \cite{bren99prl, kwia90pra,
ouzo90prl, bren91prl, kwia93pra, taps94prl, titt98prl,
titt00prl, rib00pra, thew04prl, marc04prl, sala08prl,
sala08nat} is to add an extra assumption, the
fact that a photon is detected at a specific time is
independent of the local experiment performed on that photon
\cite{aert01prl, fran09pra}. Therefore, even assuming ideal
devices, Franson's set up can only rule out local LHV theories
when this extra property is assumed. However, without it, the
results of all previous tests of the Bell-CHSH inequality using
Franson's setup can be reproduced with a local HV model like
those given in Refs.~\cite{aert01prl, cabe09prl}.

Three different strategies have been proposed to avoid the
postselection loophole in Bell tests with energy-time
entanglement:

Aerts {\em et al.} \cite{aert01prl} showed that general local
HV models are not possible using Franson's setup when very fast
local switching is used and a specific three-setting Bell
inequality (instead of the standard two-setting Bell-CHSH
inequality) is violated. This approach has two problems: It
requires a very difficult to achieve fast switching, and a
violation which is very close to the maximum violation
predicted by QM (i.e., it requires nearly perfect visibility).

Brendel {\em et al.} \cite{bren99prl} proposed replacing the
two beam splitters which are closer to the source in Franson's
setup, by switchers synchronized with the source. However,
these active switchers do not exist for photonic sources, thus
in actual experiments they are are replaced by passive beam
splitters (see, e.g., \cite{bren99prl}), so the resulting setup
suffer from the same problem of the original Franson's setup.
Recently, it has been pointed out that active switchers could
be feasible if photons are replaced by molecules
\cite{gnei08prl}.

More recently, some of us have proposed a modification of
Franson's scheme in which both the short path of the first
(second) photon and the long path of the second (first) photon
ends in the same observer \cite{cabe09prl}. In this scheme, the
rejection of events is local (i.e., it does not require
communication between the distant observers, as in Franson's
scheme), and the selection of events is independent of the
local settings. This property is the one that makes that this
scheme do not suffer from the postselection loophole that
affect {\em all} previous Bell-CHSH experiments with
energy-time or time-bin entangled photons. This scheme has
inspired a new source of electronic entanglement
\cite{frus09prb} and has been theoretically extended to
multiparty Bell experiments with energy-time entangled photons
\cite{vall09pra}.

In this Letter we present the first postselection loophole-free
violation of the the Bell-CHSH inequality using energy-time
entangled photons. The experiment is based on the scheme
proposed in \cite{cabe09prl}.

%%%%%%%%%%%%%%%%%%%%%%%%%%%%%%%%%%%%%%%%%%%%%%%%%%%%%%%%%%%%%%%%%%%

{\em Experimental setup.---}The experimental setup used for
testing the Bell-CHSH inequality with energy-time entangled
photons is schematically shown in Fig.~\ref{Fig:Setup}. A
single-longitudinal-mode laser operating at 266~nm and with an
average power of 70~mW is focused into a 5-mm-long
$\beta$-barium borate (BBO) crystal cut for type-I parametric
down-conversion luminescence \cite{hong84pra, burn70prl}. The
generated energy-time correlated photons are collimated by a
$15$~cm focal length lens put at focal distance from the
crystal, and then sent through two symmetrical interferometers.
These interferometers are unbalanced, thus one can refer to
their arms as short ($S$) and long ($L$). The optical paths of
the down-converted photons are such that coincidences counts
between detectors $D_{A}$ and $D_{B}$ can be seen only when
they propagate through the short-short or long-long two-photon
paths. These detectors are composed of interference filters
with small bandwidths ($5$~nm FWHM), single-mode optical fibers
and pigtailed avalanche photo-counting modules, that are
connected to a circuit used to record the singles and the
coincidences counts. The phases of local measurements ($\phi_A$
and $\phi_B$) are set by moving piezoelectric driven stages on
which the beam splitters are placed as it is shown in
Fig.~\ref{Fig:Setup}. The output $D_{1A}$ ($D_{2A}$)
corresponds to the projection onto
$\ket{\phi_A}=\frac{1}{\sqrt2}(\ket L+e^{i\phi_{A}}\ket{S})$
[$\ket{\phi^\perp_A}=\frac{1}{\sqrt2}(\ket
L-e^{i\phi_{A}}\ket{S})$]. A similar relation holds for the
$D_B$ detection.

The four probabilities of coincidence detection after the
interferometers can be easily calculated \cite{horn89prl}. They
depend of the initial two-photon state, $\rho$, and in the case
of down-converted photons belonging to the Bell state
$\rho=\ket{\Phi^+}\bra{\Phi^+}$, where
$\ket{\Phi^+}\equiv\frac{1}{\sqrt{2}}\left(\ket{SS}+\ket{LL}\right)$,
these probabilities will be given by
\begin{equation}
P_{ij}(\phi_A,\phi_B) = \frac{1}{4}\left[1
-(-1)^{\delta_{ij}}V_{ij} \cos\left(\phi_A+\phi_B\right)\right],
\end{equation}
%\begin{eqnarray}
%P_{11}(D_{a1},D_{b1}|\phi_A,\phi_B) &= &\frac{\eta^2}{4}\left[1
%+V_{11} \cos\left(\phi_A+\phi_B\right)\right] \nonumber \\
%P_{12}(D_{a1},D_{b2}|\phi_A,\phi_B) &= &\frac{\eta^2}{4}\left[1 -
%V_{12} \cos\left(\phi_A+\phi_B\right)\right] \nonumber \\
%P_{21}(D_{a2},D_{b1}|\phi_A,\phi_B)&= &\frac{\eta^2}{4}\left[1 -
%V_{21} \cos\left(\phi_A+\phi_B\right)\right] \nonumber \\
%P_{22}(D_{a2},D_{b2}|\phi_A,\phi_B) &= &\frac{\eta^2}{4}\left[1 +
%V_{22} \cos\left(\phi_A+\phi_B\right)\right],
%\end{eqnarray}
where $P_{ij}(\phi_A,\phi_B)$ is the probability of coincidence
detection between $D_{iA}$ and $D_{jB}$, and
%$\eta$ is the overall detectors efficiency,
$V_{ij}$ is the corresponding visibility of
the interference.

A Bell inequality can be tested by considering the coincidence
counts recorded by two detectors placed behind of the
interferometers. Nevertheless, in this case the experiment
requires an extra fair sampling assumption (e.g., that
detected photons are a statistically representative
subsensemble of the photons detected in the case where four detectors
are present), and reduces the credibility of the conclusion.
This extra assumption is not necessary in a test of the
Bell-CHSH inequality, where the coincidence counts between all
the four output ports of these interferometers are considered.

In our experiment, we recorded the coincidence counts between
the detector $D_{1A}$ ($D_{2A}$) and the detectors $D_{1B}$ and
$D_{2B}$ illustrated in Fig.~\ref{Fig:Setup}. We used these
coincidences to test the Bell-CHSH inequality and to
reconstruct the density operator of the energy-time entangled
two-photon state used in the test of the Bell-CHSH inequality.

Our experiment is not free of the locality loophole, since the
distance between Alice and Bob's local experiments is around
one meter, which is not enough to prevent reciprocal influences
between the two local phase settings. Furthermore, our
experiment, like any other photonic Bell experiment, is also
not free of the detection loophole, since the overall detection
efficiency is less than $15\%$. The main aim of the experiment
is to provide a proof of principle that, with enough separation
and more efficient detectors, a conclusive violation of the
Bell-CHSH inequality can be observed with energy-time entangled
photons.

Any local HV model should satisfy the Bell-CHSH-inequality
\begin{equation}
S \leq 2,
\end{equation}
where
\begin{multline}
S \equiv E(\phi_A,\phi_B) + E(\phi_A',\phi_B)\\
+E(\phi_A,\phi_B') - E(\phi_A',\phi_B')\,,
\end{multline}
and
\begin{multline}
E(\phi_A,\phi_B) \equiv P_{11}(\phi_A,\phi_B)
+ P_{22}(\phi_A,\phi_B) \\
- P_{12}(\phi_A,\phi_B)- P_{21}(\phi_A,\phi_B).
%{P_{11}(\phi_A,\phi_B)
%+ P_{22}(\phi_A,\phi_B) +
%P_{12}(\phi_A,\phi_B) + P_{21}(\phi_A,\phi_B)}.
\label{E}
\end{multline}

If all the four possible interference curves have the same
average visibility $V$, the initial two-photon state is
$\rho=\ket{\Phi^+}\bra{\Phi^+}$, and the phase settings are
\begin{equation}
 \phi_A = \frac{\pi}{4}, \,\,\,\,\,\, \phi_B = 0, \,\,\,\,\,\, \phi_A' = -
 \frac{\pi}{4},\,\,\,\,\,\, \phi_B' = \frac{\pi}{2},
 \label{phases}
\end{equation}
then, the expected value for $S$ is $S = 2\sqrt{2}V$.
Therefore, we will have an experimental violation of $S$
whenever the average visibility of the two-photon interferences
is larger than $1/\sqrt{2} \approx 0.71$.

\begin{figure}[t]
\vspace{-1.5cm}
\begin{center}
\includegraphics[width=9cm]{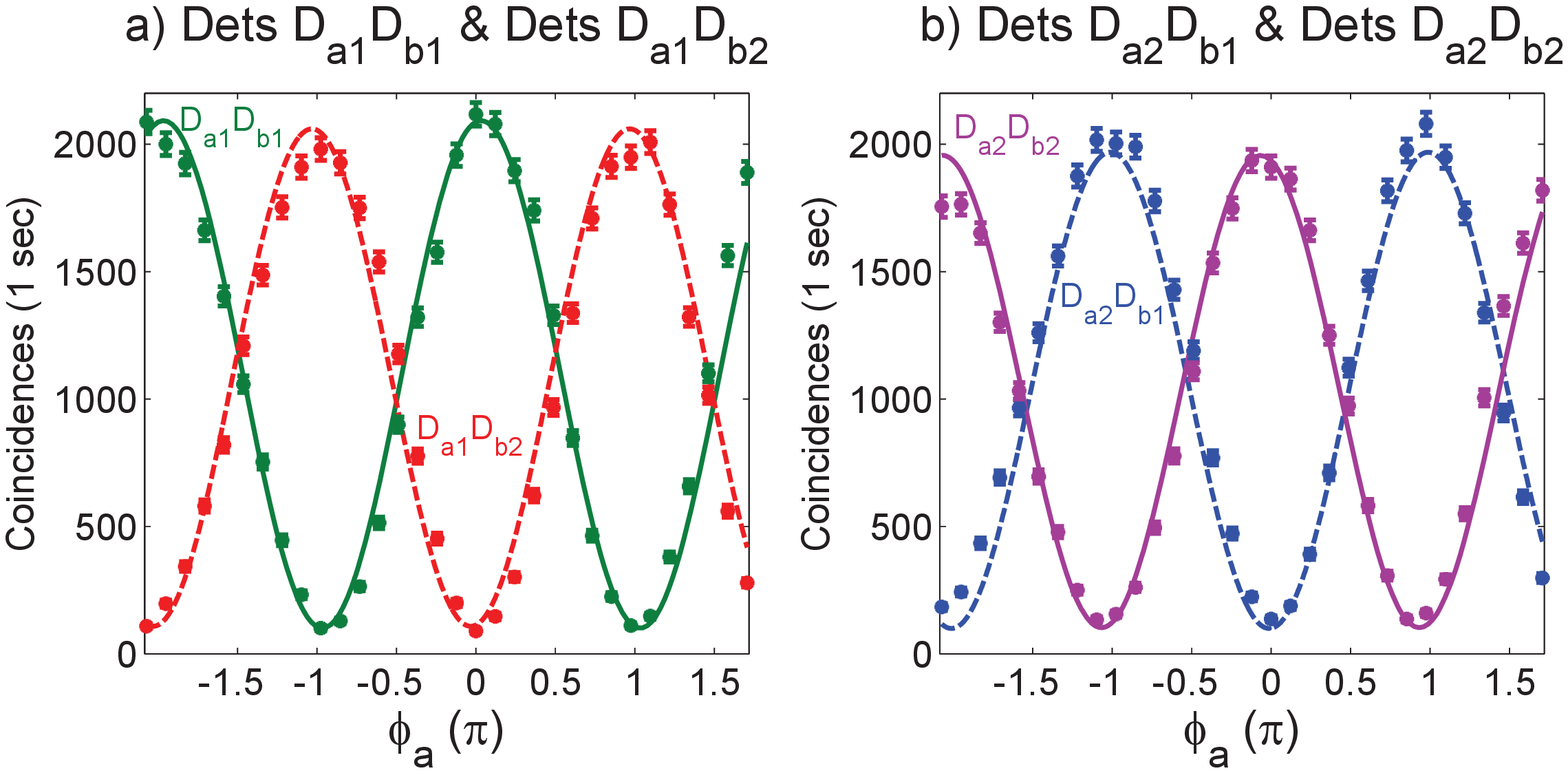}
\end{center}
\vspace{-2cm} \caption{The four coincidence curves obtained while
$\phi_B=0$.} \label{Fig:phi0}
\end{figure}

\begin{figure}[t]
\vspace{-1.5cm}
\begin{center}
\includegraphics[width=9cm]{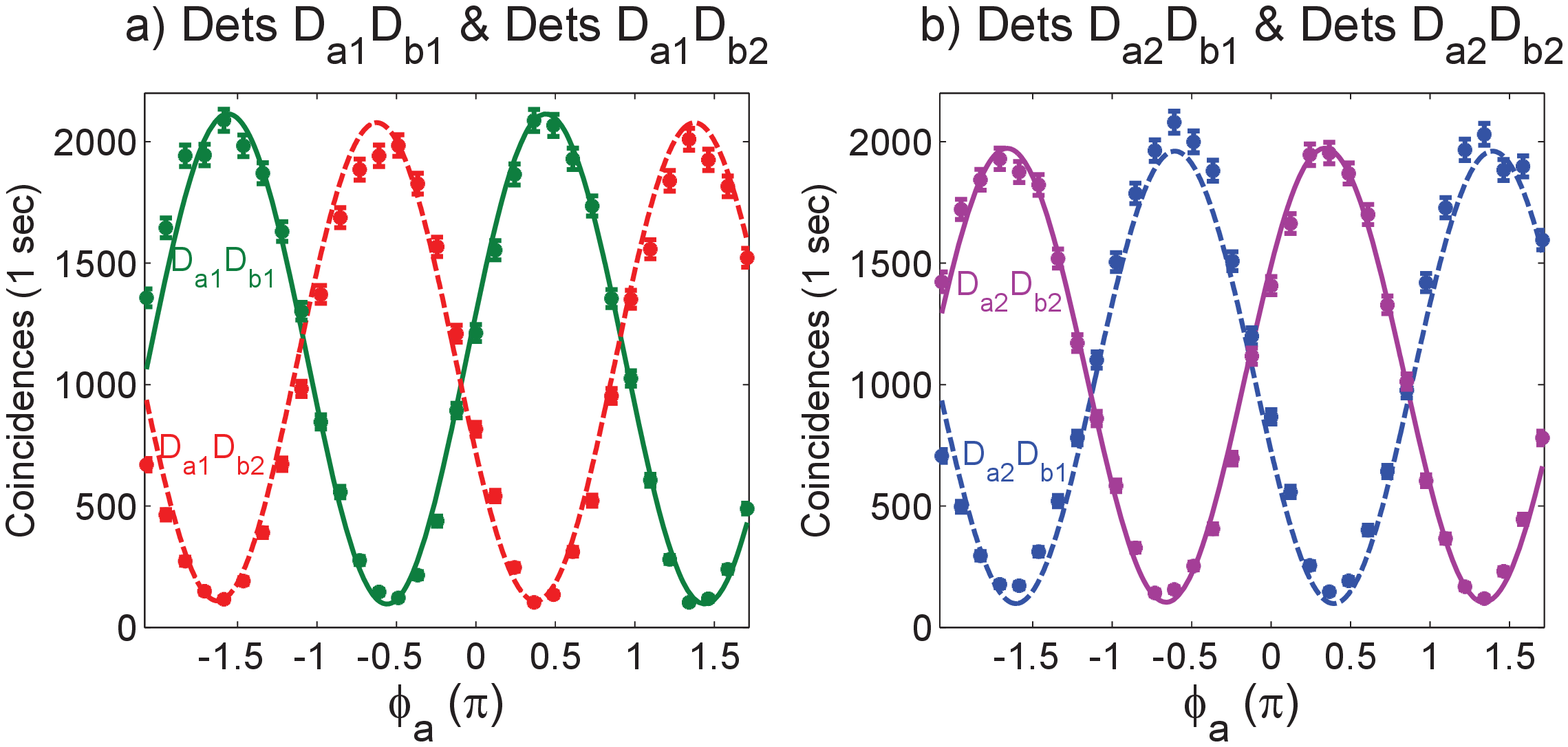}
\end{center}
\vspace{-2cm} \caption{The four coincidence curves obtained while
$\phi_B=\frac{\pi}{2}$.} \label{Fig:phipi2}
\end{figure}

The correlation functions $E$ that appear in the Bell-CHSH
inequality can be determined directly from the coincidence
counts recorded by the four detectors. The experimental
coincidence counts recorded between the four detectors, when
$\phi_B=0$ and $\phi_B=\frac{\pi}{2}$, are shown in
Fig.~\ref{Fig:phi0} and Fig.~\ref{Fig:phipi2}, respectively.
From these curves we calculated the values of the probability
correlation functions $E$ and, therefore, $S$. We calculated
$S$ in two different ways: directly from the experimental
points obtained, and from the values of the fit of the
experimental data. Both are shown in Table~\ref{tab1}.

The mean visibility observed in the curves of
Figs.~\ref{Fig:phi0} and \ref{Fig:phipi2} is $V=0.90\pm0.015$,
and therefore the expected value of $S$ is $S_{\rm
exp}=2\sqrt2V=2.54\pm0.04$. The experimental results obtained
for $S$ (see Table~\ref{tab1}) are slightly lower than $S_{\rm
exp}$ due to small phase-shifts that exists between the eight
curves recorded. However, they correspond to a violation of the
Bell-CHSH inequality by 20 standard deviations.

%%%%%%%%%%%%%%%%%%%%%%%%%%%%%%%%%%%%%%%%%%%%%%%%%%%%%%%%%%%%%%%%%%%
% Fig. 2
%%%%%%%%%%%%%%%%%%%%%%%%%%%%%%%%%%%%%%%%%%%%%%%%%%%%%%%%%%%%%%%%%%%

%%%%%%%%%%%%%%%%%%%%%%%%%%%%%%%%%%%%%%%%%%%%%%%%%%%%%%%%%%%%%%%%%%%
% Fig. 3
%%%%%%%%%%%%%%%%%%%%%%%%%%%%%%%%%%%%%%%%%%%%%%%%%%%%%%%%%%%%%%%%%%%

%%%%%%%%%%%%%%%%%%%%%%%%%%%%%%%%%%%%%%%%%%%%%%%%%%%%%%%%%%%%%%%%%%%
% Table I
%%%%%%%%%%%%%%%%%%%%%%%%%%%%%%%%%%%%%%%%%%%%%%%%%%%%%%%%%%%%%%%%%%%

\begin{table}[t]
\begin{ruledtabular}
\begin{tabular}{cccc}
&$S$ &$\delta S$ &Std dev.  \\
\hline
(a) &$2.468$ &$0.024$ &$19.57$ \\
(b) &$2.473$ &$0.024$ &$20.02$
\end{tabular}
\end{ruledtabular}
\caption{Experimental violation of the Bell-CHSH inequality
obtained (a) from the experimental points recorded, and (b)
based on the experimental fits. In both cases we dis not
subtract accidental coincidences.} \label{tab1}
\end{table}

%%%%%%%%%%%%%%%%%%%%%%%%%%%%%%%%%%%%%%%%%%%%%%%%%%%%%%%%%%%%%%%%%%%

The energy-time two photon state was tested by quantum
tomography. The reconstructed density operator was obtained
considering an analogy between the projective
phase-measurements and the measurements used for doing the
standard quantum tomography of polarization two-photon states
\cite{jame01pra}. The data acquired and the analogy between
these measurements are shown in Table~\ref{tab2}.

%%%%%%%%%%%%%%%%%%%%%%%%%%%%%%%%%%%%%%%%%%%%%%%%%%%%%%%%%%%%%%%%%%%
% Table II
%%%%%%%%%%%%%%%%%%%%%%%%%%%%%%%%%%%%%%%%%%%%%%%%%%%%%%%%%%%%%%%%%%%

\begin{center}
\begin{table}
\vspace{0.5cm}
\begin{tabular}{c|c|c|c|c} \hline \hline
\multicolumn{5}{c}{\textbf{Reconstructing $\rho$}} \\
\hline \hline Mea. &Pol. proj. &Prop. Mode A &Prop. Mode B & $C_{T}$
\\ \cline{1-5}
1 & $\left\vert VV\right\rangle \left\langle VV\right\vert $ &\ket L &\ket L &3058 \\
2 & $\left\vert HV\right\rangle \left\langle HV\right\vert $ &\ket S &\ket L &31 \\
3 & $\left\vert HH\right\rangle \left\langle HH\right\vert $ &\ket S &\ket S &3416 \\
4 & $\left\vert VH\right\rangle \left\langle VH\right\vert $ &\ket L &\ket S &35 \\
5 & $\left\vert Vl\right\rangle \left\langle Vl\right\vert $ &\ket L &$\ket{\phi_B=\frac{\pi}{2}}$ &1737 \\
6 & $\left\vert Hl\right\rangle \left\langle Hl\right\vert $ &\ket S &$\ket{\phi_B=\frac{\pi}{2}}$ &1799 \\
7 & $\left\vert HD\right\rangle \left\langle HD\right\vert $ &\ket S &$\ket{\phi_B=0}$ &1708 \\
8 & $\left\vert VD\right\rangle \left\langle VD\right\vert $ &\ket L &$\ket{\phi_B=0}$ &1797 \\
9 & $\left\vert lD\right\rangle \left\langle lD\right\vert $ &${\ket{\phi_A=\frac{\pi}{2}}}$ &$\ket{\phi_B=0}$ &1795 \\
10 & $\left\vert DD\right\rangle \left\langle DD\right\vert $ &$\ket{\phi_A=0}$ &$\ket{\phi_B=0}$ &3304 \\
11 & $\left\vert Dl\right\rangle \left\langle DL\right\vert $ &$\ket{\phi_A=0}$ &$\ket{\phi_B=\frac{\pi}{2}}$ &1727 \\
12 & $\left\vert DV\right\rangle \left\langle DV\right\vert $ &$\ket{\phi_A=0}$ &\ket L &1762 \\
13 & $\left\vert DH\right\rangle \left\langle DH\right\vert $ &$\ket{\phi_A=0}$ &\ket S &1801 \\
14 & $\left\vert rH\right\rangle \left\langle rH\right\vert $ &${\ket{\phi_A=\frac{\pi}{2}}}$ &\ket S &1713 \\
15 & $\left\vert rV\right\rangle \left\langle rV\right\vert $ &${\ket{\phi_A=\frac{\pi}{2}}}$ &\ket L &1744 \\
16 & $\left\vert rl\right\rangle \left\langle rl\right\vert$
&${\ket{\phi_A=\frac{\pi}{2}}}$ &$\ket{\phi_B=\frac{\pi}{2}}$ &97
\end{tabular}
\caption{Summary of the phase-measurements used to perform the
reconstruction of the energy-time entangled photon state. We
compare the measurements with those used in the quantum
tomography of two polarized qubits. $\ket{H}$, $\ket{V}$,
$\ket{D}$, $\ket{l}$ and $\ket{r}$ represent polarized qubits
with horizontal, vertical, diagonal, left- and right-circular
polarization, respectively. $\ket{S}$ and $\ket{L}$ represent
short and long path states. The total coincidence, $C_{T}$, was
recorded in two seconds.} \label{tab2}
\end{table}
\end{center}

%%%%%%%%%%%%%%%%%%%%%%%%%%%%%%%%%%%%%%%%%%%%%%%%%%%%%%%%%%%%%%%%%%%

\begin{figure}[t]
\vspace{-1.5cm}
\centering
\includegraphics[width=9cm]{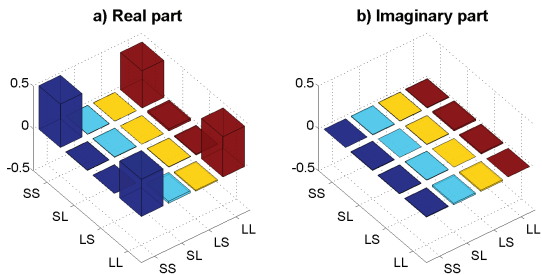}
\vspace{-2.0cm}
\caption{Tomographic reconstruction of the quantum state used for the Bell-CHSH test.}
\label{Fig:rho}
\end{figure}

The density operator $\rho_{\rm exp}$ obtained after numerical
optimization is shown in Fig.~\ref{Fig:rho}. It has a fidelity
\cite{jozs94jmo} of $0.93\pm0.02$ with the Bell state
$\rho=\ket{\Phi^+}\bra{\Phi^+}$. For this reconstruction, we
did not consider accidental substraction. In this case, the
predicted maximum value of $S$ is $\text{tr}(S\rho_{\rm
exp})=2.488$. The actual measured values of $S$ (see
Table~\ref{tab1}) are very close to this value. By considering
accidental subtraction, the fidelity of the reconstructed state
was $0.97\pm0.02$.

%%%%%%%%%%%%%%%%%%%%%%%%%%%%%%%%%%%%%%%%%%%%%%%%%%%%%%%%%%%%%%%%%%%

{\em Conclusions.---}We have presented the first violation of
the Bell-CHSH inequality using energy-time entangled photons
which is free of the postselection loophole reported by Aerts
{\em et al.} \cite{aert01prl} and which affects all previous
Bell experiments using energy-time or time-bin entangled
photons \cite{kwia90pra, ouzo90prl, bren91prl, kwia93pra,
taps94prl, titt98prl, titt00prl, rib00pra, thew04prl,
marc04prl, sala08prl, sala08nat}.

%%%%%%%%%%%%%%%%%%%%%%%%%%%%%%%%%%%%%%%%%%%%%%%%%%%%%%%%%%%%%%%%%%%
% Fig. 4
%%%%%%%%%%%%%%%%%%%%%%%%%%%%%%%%%%%%%%%%%%%%%%%%%%%%%%%%%%%%%%%%%%%

%%%%%%%%%%%%%%%%%%%%%%%%%%%%%%%%%%%%%%%%%%%%%%%%%%%%%%%%%%%%%%%%%%%

The experiment is not free of the locality and detection
loopholes. Increasing the length of the interferometers and
adding a fast switch between the local setting is required to
scape from the locality loophole. The required values and the
stability problems were discussed in \cite{cabe09prl}.
Increasing the efficiency of the photodetectors or replacing
low-energy photons by easily detectable particles, like massive
particles, is required to escape from the detection loophole.
However, the experiment provides a proof of principle that the
scheme proposed in \cite{cabe09prl} is suitable for a testing
the Bell-CHSH inequality with energy-time entangled particles.
Further developments might include nonphotonic versions of the
setup and larger-scale implementations.

%%%%%%%%%%%%%%%%%%%%%%%%%%%%%%%%%%%%%%%%%%%%%%%%%%%%%%%%%%%%%%%%%%%

\begin{acknowledgments}
GL acknowledges support by the FONDECYT Project No. 11085055.
AC acknowledges support by the Spanish MCI Project No.
FIS2008-05596, and the Junta de Andaluc\'{\i}a Excellence
Project No. P06-FQM-02243.
\end{acknowledgments}

%%%%%%%%%%%%%%%%%%%%%%%%%%%%%%%%%%%%%%%%%%%%%%%%%%%%%%%%%%%%%%%%%%%

%\bibliography{../../../bibliografia/qi-bibliografia}% Produces the bibliography via BibTeX.

\begin{thebibliography}{10}
\expandafter\ifx\csname natexlab\endcsname\relax\def\natexlab#1{#1}\fi
\expandafter\ifx\csname bibnamefont\endcsname\relax
 \def\bibnamefont#1{#1}\fi
\expandafter\ifx\csname bibfnamefont\endcsname\relax
 \def\bibfnamefont#1{#1}\fi
\expandafter\ifx\csname citenamefont\endcsname\relax
 \def\citenamefont#1{#1}\fi
\expandafter\ifx\csname url\endcsname\relax
 \def\url#1{\texttt{#1}}\fi
\expandafter\ifx\csname urlprefix\endcsname\relax\def\urlprefix{URL }\fi
\providecommand{\bibinfo}[2]{#2}
\providecommand{\eprint}[2][]{\url{#2}}

\bibitem[{\citenamefont{Bell}(1964)}]{bell64phy}
  \bibinfo{author}{\bibfnamefont{J.~S.} \bibnamefont{Bell}},
 \bibinfo{journal}{Physics (Long Island City, N.Y.)} \textbf{\bibinfo{volume}{1}}, \bibinfo{pages}{195}
 (\bibinfo{year}{1964}).

\bibitem[{\citenamefont{Clauser et~al.}(1969)\citenamefont{Clauser, Horne,
 Shimony, and Holt}}]{clau69prl}
\bibinfo{author}{\bibfnamefont{J.~F.} \bibnamefont{Clauser}},
 \bibinfo{author}{\bibfnamefont{M.~A.} \bibnamefont{Horne}},
 \bibinfo{author}{\bibfnamefont{A.}~\bibnamefont{Shimony}}, \bibnamefont{and}
 \bibinfo{author}{\bibfnamefont{R.~A.} \bibnamefont{Holt}},
 \bibinfo{journal}{Phys. Rev. Lett.} \textbf{\bibinfo{volume}{23}},
 \bibinfo{pages}{880} (\bibinfo{year}{1969}).

%%%%%%%%%%%%%%%%%%%%%%%%%%%%%%%%%%%%%%%%%%%%%%%%%%%%%%%%%%%%%%%%%%%

\bibitem{barr05prl}
 J. Barrett, L. Hardy, and A. Kent
 %``No signalling and quantum key distribution'',
 Phys. Rev. Lett. {\bf 95}, 010503 (2005).

\bibitem{buhr09rmp}
 H. Buhrman, R. Cleve, S. Massar, and R. de Wolf,
 %``Non-locality and communication complexity'',
 Rev. Mod. Phys.; \eprint{arXiv:0907.3584}.

\bibitem{piro09nat}
S. Pironio {\em et al.},
% S. Pironio, A. Ac\'{\i}n, S. Massar, A. Boyer de la Giroday,
% D. N. Matsukevich, P. Maunz, S. Olmschenk, D. Hayes, L. Luo,
% T. A. Manning, and C. Monroe,
%``Random numbers certified by Bell's theorem'',
 \eprint{arXiv:0911.3427}.

%%%%%%%%%%%%%%%%%%%%%%%%%%%%%%%%%%%%%%%%%%%%%%%%%%%%%%%%%%%%%%%%%%%

\bibitem{free72prl}
 S. J. Freedman and J. F. Clauser,
%``Experimental test of local hidden-variable theories'',
 Phys. Rev. Lett. {\bf 28}, 938 (1972).

\bibitem{aspe82prl}
 A. Aspect, J. Dalibard, and G. Roger,
%``Experimental test of Bell's inequalities using time-varying
%analyzers'',
 Phys. Rev. Lett. {\bf 49}, 1804 (1982).

\bibitem{weih98prl}
 G. Weihs, T. Jennewein, C. Simon, H. Weinfurter, and A. Zeilinger,
%``Violation of Bell's inequality under strict
%Einstein locality conditions'',
 Phys. Rev. Lett. {\bf 81}, 5039 (1998).

\bibitem{rowe01nat}
 M. A. Rowe, D. Kielpinski, V. Meyer,
 C. A. Sackett, W. M. Itano, C. Monroe, and D. J. Wineland,
%``Experimental violation of a Bell's inequality
%with efficient detection'',
 Nature (London) {\bf 409}, 791 (2001).

\bibitem{mats08prl}
 D. N. Matsukevich, P. Maunz, D. L. Moehring,
 S. Olmschenk, and C. Monroe,
%``Bell inequality violation with two remote atomic qubits'',
 Phys. Rev. Lett. {\bf 100}, 150404 (2008).

%%%%%%%%%%%%%%%%%%%%%%%%%%%%%%%%%%%%%%%%%%%%%%%%%%%%%%%%%%%%%%%%%%%


\bibitem{aspe99nat}
 A. Aspect,
%``Bell's inequality test: More ideal than ever'',
 Nature (London) {\bf 398}, 189 (1999).

%%%%%%%%%%%%%%%%%%%%%%%%%%%%%%%%%%%%%%%%%%%%%%%%%%%%%%%%%%%%%%%%%%%

\bibitem[{\citenamefont{Santos}(1991)}]{sant91prl}
\bibinfo{author}{\bibfnamefont{E.}~\bibnamefont{Santos}},
  \bibinfo{journal}{Phys. Rev. Lett.} \textbf{\bibinfo{volume}{66}},
  \bibinfo{pages}{1388} (\bibinfo{year}{1991}).

%%%%%%%%%%%%%%%%%%%%%%%%%%%%%%%%%%%%%%%%%%%%%%%%%%%%%%%%%%%%%%%%%%%

\bibitem{fran89prl}
 J. D. Franson,
%``Bell inequality for position and time'',
 Phys. Rev. Lett. {\bf 62}, 2205 (1989).

%%%%%%%%%%%%%%%%%%%%%%%%%%%%%%%%%%%%%%%%%%%%%%%%%%%%%%%%%%%%%%%%%%%

\bibitem{bren99prl}
 J. Brendel, N. Gisin, W. Tittel, and H. Zbinden,
%``Pulsed energy-time entangled twin-photon source for quantum communication'',
 Phys. Rev. Lett. {\bf 82}, 2594 (1999).

%%%%%%%%%%%%%%%%%%%%%%%%%%%%%%%%%%%%%%%%%%%%%%%%%%%%%%%%%%%%%%%%%%%

\bibitem{kwia90pra}
 P. G. Kwiat, W. A. Vareka, C. K. Hong, H. Nathel, and R. Y. Chiao,
 %``Correlated two-photon interference in a dual-beam Michelson interferometer'',
 Phys. Rev. A {\bf 41}, 2910 (1990).

\bibitem{ouzo90prl}
 Z. Y. Ou, X. Y. Zou, L. J. Wang, and L. Mandel,
%``Observation of nonlocal interference in separated photon channels'',
 Phys. Rev. Lett. {\bf 65}, 321 (1990).

\bibitem{bren91prl}
 J. Brendel, E. Mohler, and W. Martienssen,
%``Time-resolved dual-beam two-photon interferences with high visibility'',
 Phys. Rev. Lett. {\bf 66}, 1142 (1991).

\bibitem{kwia93pra}
 P. G. Kwiat, A. M. Steinberg, and R. Y. Chiao,
%``High-visibility interference in a Bell-inequality experiment for energy and time'',
 Phys. Rev. A {\bf 47}, R2472 (1993).

\bibitem{taps94prl}
 P. R. Tapster, J. G. Rarity, and P. C. M. Owens,
%``Violation of Bell's inequality over 4 km of optical fiber'',
 Phys. Rev. Lett. {\bf 73}, 1923 (1994).

\bibitem{titt98prl}
 W. Tittel, J. Brendel, H. Zbinden, and N. Gisin,
%``Violation of Bell inequalities by photons more than 10 km apart'',
 Phys. Rev. Lett. {\bf 81}, 3563 (1998).

\bibitem{titt00prl}
 W. Tittel, J. Brendel, H. Zbinden, and N. Gisin,
%``Quantum cryptography using entangled photons in energy-time Bell states'',
 Phys. Rev. Lett. {\bf 84}, 4737 (2000).

\bibitem{rib00pra}
 G. Ribordy, J. Brendel, J-D. Gautier, N. Gisin, and H. Zbinden,
%``Long-distance entanglement-based quantum key distribution'',
 Phys. Rev. A {\bf 63}, 012309 (2000).

\bibitem{thew04prl}
 R. T. Thew, A. Ac\'{\i}n, H. Zbinden, and N. Gisin,
%``A Bell-type test of energy-time entangled qutrits'',
Phys. Rev. Lett. {\bf 93}, 010503 (2004).

\bibitem{marc04prl}
 I. Marcikic, H. de Riedmatten, W. Tittel, H. Zbinden, M. Legr\'{e},
 and N. Gisin,
%``Distribution of time-bin entangled qubits over 50 km
%of optical fiber'',
 Phys. Rev. Lett. {\bf 93}, 180502 (2004).

\bibitem{sala08prl}
 D. Salart, A. Baas, J. A. W. van Houwelingen, N. Gisin, and H. Zbinden,
%``Space-like separation in a Bell test assuming gravitationally induced collapses'',
 Phys. Rev. Lett. {\bf 100}, 220404 (2008).

\bibitem{sala08nat}
 D. Salart, A. Baas, C. Branciard, N. Gisin, and H. Zbinden,
 %``Testing the speed of 'spooky action at a distance'''
 Nature (London) {\bf 454}, 861 (2008).

%%%%%%%%%%%%%%%%%%%%%%%%%%%%%%%%%%%%%%%%%%%%%%%%%%%%%%%%%%%%%%%%%%%

\bibitem{aert01prl}
 S. Aerts, P. G. Kwiat, J.-\AA. Larsson, and M. \.{Z}ukowski,
%``Two-photon Franson-type experiments and local realism'',
 Phys. Rev. Lett. {\bf 83}, 2872 (1999); {\bf 86}, 1909 (2001).

%%%%%%%%%%%%%%%%%%%%%%%%%%%%%%%%%%%%%%%%%%%%%%%%%%%%%%%%%%%%%%%%%%%

\bibitem[{\citenamefont{Cabello et~al.}(2009)\citenamefont{Cabello, Rossi,
 Vallone, {De Martini}, and Mataloni}}]{cabe09prl}
\bibinfo{author}{\bibfnamefont{A.}~\bibnamefont{Cabello}},
 \bibinfo{author}{\bibfnamefont{A.}~\bibnamefont{Rossi}},
 \bibinfo{author}{\bibfnamefont{G.}~\bibnamefont{Vallone}},
 \bibinfo{author}{\bibfnamefont{F.}~\bibnamefont{{De Martini}}},
 \bibnamefont{and} \bibinfo{author}{\bibfnamefont{P.}~\bibnamefont{Mataloni}},
 \bibinfo{journal}{Phys. Rev. Lett.} \textbf{\bibinfo{volume}{102}},
 \bibinfo{pages}{040401} (\bibinfo{year}{2009}).

\bibitem{fran09pra}  
 J. D. Franson,
%``Nonclassical nature of dispersion cancellation and nonlocal interferometry'',
 Phys. Rev. A {\bf 80}, 032119 (2009).

%%%%%%%%%%%%%%%%%%%%%%%%%%%%%%%%%%%%%%%%%%%%%%%%%%%%%%%%%%%%%%%%%%%

\bibitem{gnei08prl}
 C. Gneiting and K. Hornberger,
%``Bell test for the free motion of material particles'',
 Phys. Rev. Lett. {\bf 101}, 260503 (2008).

%%%%%%%%%%%%%%%%%%%%%%%%%%%%%%%%%%%%%%%%%%%%%%%%%%%%%%%%%%%%%%%%%%%

\bibitem{frus09prb}
 D. Frustaglia and A. Cabello,
%``Electronic entanglement via quantum Hall interferometry in analogy to an optical method'',
 Phys. Rev. B {\bf 80}, 201312(R) (2009).

\bibitem{vall09pra}
 G. Vallone, P. Mataloni, and A. Cabello,
%``Multiparty multilevel energy-time entanglement''
 \eprint{arXiv:0912.4419}

%%%%%%%%%%%%%%%%%%%%%%%%%%%%%%%%%%%%%%%%%%%%%%%%%%%%%%%%%%%%%%%%%%%

\bibitem[{\citenamefont{Hong and Mandel}(1984)}]{hong84pra}
\bibinfo{author}{\bibfnamefont{C.~K.} \bibnamefont{Hong}} \bibnamefont{and}
 \bibinfo{author}{\bibfnamefont{L.}~\bibnamefont{Mandel}},
 \bibinfo{journal}{Phys. Rev. A} \textbf{\bibinfo{volume}{31}},
 \bibinfo{pages}{2409} (\bibinfo{year}{1984}).

\bibitem[{\citenamefont{Burnham and Weinberg}(1970)}]{burn70prl}
\bibinfo{author}{\bibfnamefont{D.~C.} \bibnamefont{Burnham}} \bibnamefont{and}
 \bibinfo{author}{\bibfnamefont{D.~L.} \bibnamefont{Weinberg}},
 \bibinfo{journal}{Phys. Rev. Lett.} \textbf{\bibinfo{volume}{25}},
 \bibinfo{pages}{84} (\bibinfo{year}{1970}).

\bibitem[{\citenamefont{Horne et~al.}(1989)\citenamefont{Horne, Shimony, and
 Zeilinger}}]{horn89prl}
\bibinfo{author}{\bibfnamefont{M.~A.} \bibnamefont{Horne}},
 \bibinfo{author}{\bibfnamefont{A.}~\bibnamefont{Shimony}}, \bibnamefont{and}
 \bibinfo{author}{\bibfnamefont{A.}~\bibnamefont{Zeilinger}},
 \bibinfo{journal}{Phys. Rev. Lett.} \textbf{\bibinfo{volume}{62}},
 \bibinfo{pages}{2209} (\bibinfo{year}{1989}).

%\bibitem[{\citenamefont{Cirel'son}(1980)}]{cire80lmp}
%\bibinfo{author}{\bibfnamefont{B.~S.} \bibnamefont{Cirel'son}},
% \bibinfo{journal}{Lett. Math. Phys} \textbf{\bibinfo{volume}{4}},
% \bibinfo{pages}{93} (\bibinfo{year}{1980}).

%\bibitem[{\citenamefont{Fano}(1957)}]{fano57rmp}
%\bibinfo{author}{\bibfnamefont{U.}~\bibnamefont{Fano}},
%\bibinfo{journal}{Rev.
% Mod. Phys.} \textbf{\bibinfo{volume}{29}}, \bibinfo{pages}{74}
% (\bibinfo{year}{1957}).

\bibitem[{\citenamefont{James et~al.}(2001)\citenamefont{James, Kwiat, Munro,
 and White}}]{jame01pra}
\bibinfo{author}{\bibfnamefont{D.~F.~V.} \bibnamefont{James}},
 \bibinfo{author}{\bibfnamefont{P.~G.} \bibnamefont{Kwiat}},
 \bibinfo{author}{\bibfnamefont{W.~J.} \bibnamefont{Munro}}, \bibnamefont{and}
 \bibinfo{author}{\bibfnamefont{A.~G.} \bibnamefont{White}},
 \bibinfo{journal}{Phys. Rev. A} \textbf{\bibinfo{volume}{64}},
 \bibinfo{pages}{052312} (\bibinfo{year}{2001}).

\bibitem[{\citenamefont{Jozsa}(1994)}]{jozs94jmo}
\bibinfo{author}{\bibfnamefont{R.}~\bibnamefont{Jozsa}}, \bibinfo{journal}{J.
 Mod. Opt.} \textbf{\bibinfo{volume}{41}}, \bibinfo{pages}{2315}
 (\bibinfo{year}{1994}).

\end{thebibliography}

%%%%%%%%%%%%%%%%%%%%%%%%%%%%%%%%%%%%%%%%%%%%%%%%%%%%%%%%%%%%%%%%%%%

\end{document}